\documentclass[9pt,letterpaper,english,reprint]{revtex4-1}
\usepackage[T1]{fontenc}
\usepackage[latin9]{inputenc}
\usepackage{amsmath}
\usepackage{amssymb}
\usepackage{graphicx}
\usepackage{esint}

\usepackage{indentfirst}
\usepackage{wasysym}
\usepackage{dsfont}
\usepackage{subcaption}

\usepackage{eqnarray}

\begin{document}

\title{Calibration of wavefront distortion in light modulator setup by Fourier
analysis of multi-beam interference}

\author{Adam Leszczy\'nski}
\affiliation{Institute of Experimental Physics, Faculty of Physics, University of Warsaw, Pasteura 5, 02-093 Warsaw, Poland}
\author{Wojciech Wasilewski}

\affiliation{Institute of Experimental Physics, Faculty of Physics, University of Warsaw, Pasteura 5, 02-093 Warsaw, Poland}




\begin{abstract}
We present a method to calibrate wavefront distortion of the
spatial light modulator setup by registering far field images of several
Gaussian beams diffracted off the modulator. The Fourier transform
of resulting interference images reveals phase differences between typically 5 movable points on the modulator. Repeating this measurement
yields wavefront surface. Next, the amplitude efficiency is calibrated
be registering near field image. As a verification we produced a superposition
of 7th and 8th Bessel beams with different phase velocities
and observed their interference.
\end{abstract}

\maketitle

\section{Introduction}

Phase only spatial light modulator (SLM) finds numerous applications
including optical tweezers \cite{tweezer, tweezer2,tweezer3},
microscopy \cite{mmm, spiral,isotripic}, or manipulation
of atoms \cite{bose-einstein, dynamic, key-2}. Typically in
addition to voltage-controlled phase delay the diffracted beam is subject
to wavefront distortion due to SLM curvature, wavefront error of
the illuminating beam and possibly the optics between the SLM and
final plane of application.
The SLM curvature can be calibrated using surface metrology interferometers \cite{twyman-green,zernike}.
It is more useful however to calibrate the wavefront at the output of the entire setup. 
This is possible using for example  Shack-Hartmann method \cite{schack-hartman} which measures phase profile gradient 
or using iterative Gerchberg-Saxon algorithm \cite{GS-alg_vortex, GS-alg} which is unfortunately sensitive to starting data and requires rather high dynamic range.
Recently methods which aim directly at achieving perfect focusing have been developed \cite{nature,po_nature}. 
They rely on activating two regions on the SLM and observing the interference between two light  bundles produced this way. 
This can be done on a camera \cite{po_nature} or even using fluorescent-particle \cite{nature}. 
The entire wavefront can be reconstructed by scanning one of the regions across the whole SLM.

Here we propose an alternative method for reconstructing wavefront in a particular optical setup involving SLM. 
To achieve higher throughput and noise immunity we use several regions on the SLM simultaneously and analyze resulting interference images by Fourier transform to extract phase differences.
By displaying a number of small phase gratings we diffract parallel beams which focus and interfere together on a far field camera. 
In addition we correct for nonuniform illumination of the SLM by registering the near field image of diffracted beam.

Our method works with any setup sufficient for registering both near and far field images of diffracted beams.
The implementation presented here bases entirely on singlet lenses and enables diagnosing global wavefront properties with any 
controlled-exposure digital camera placed in the far field.  We use second near field camera for an amplitude calibration but even without 
it full calibration can be accomplished. 

This paper is organized as follow. 
Section \ref{sec:setup} describes the calibration setup. 
Section \ref{sec:grating} details using phase grating to diffract arbitrary beam. 
Section \ref{sec:phase} presents a method to retrieve wavefront distortion from interference of multiple Gaussian beams in Fourier plane. 
Section \ref{sec:amplitude} details the amplitude calibration process. 
Finally section  \ref{sec:verify} describes the way of using a calibrated SLM and presents several methods to verify the calibration.

\section{Experimental Setup}

\label{sec:setup}

\begin{figure}[h]
\centering \includegraphics[width=1\linewidth]{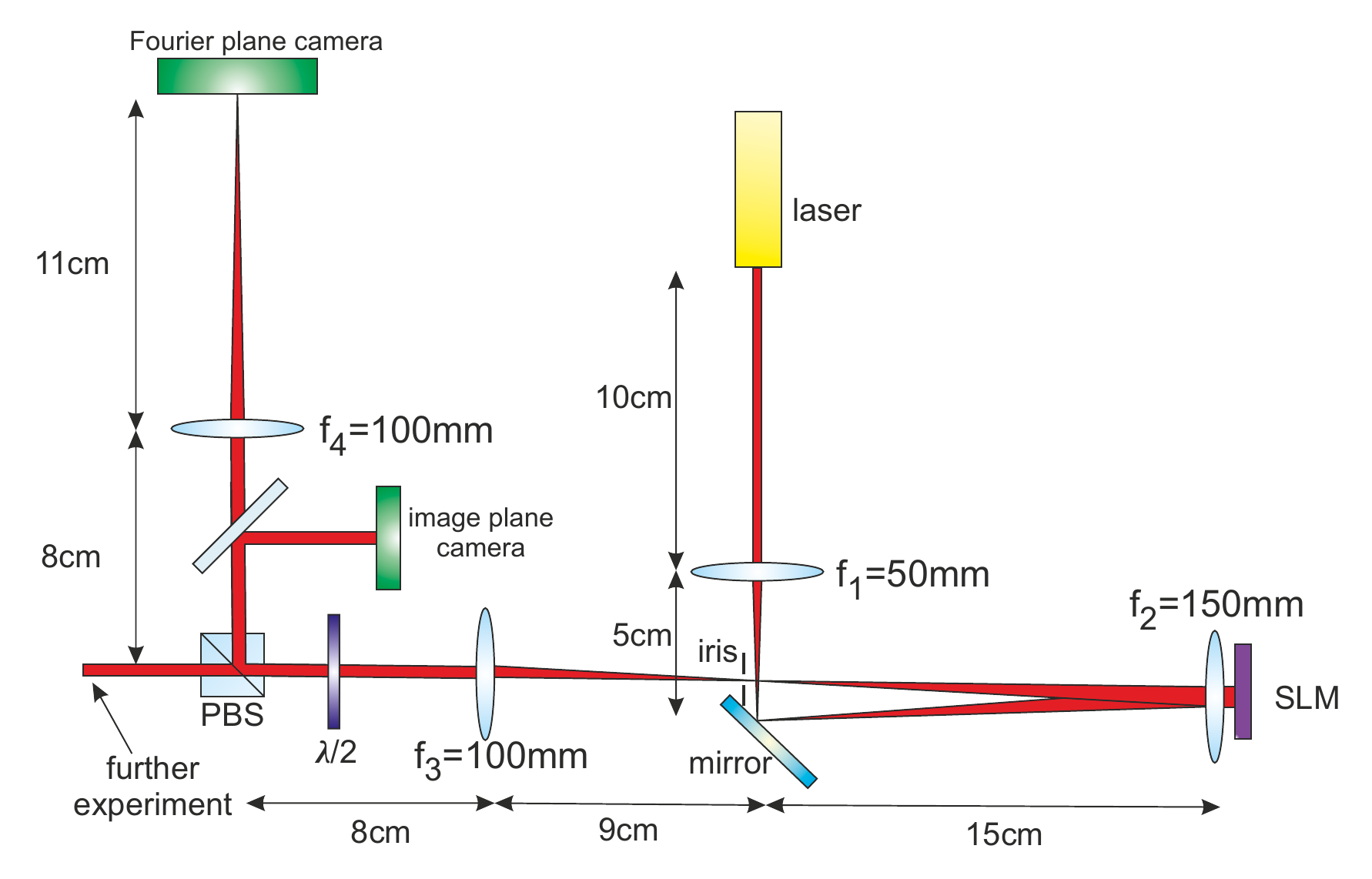} \caption{Setup for producing arbitrary beams and calibrating SLM. Lenses $\textnormal{f}_{1}$
and $\textnormal{f}_{2}$ form a telescope to illuminate entire SLM.
Iris is used to filter out only 1st order diffracted light. $\textnormal{f}_{2}$
and $\textnormal{f}_{3}$ create SLM image on image plane camera and
rely the beam to further experiment. Lens $\textnormal{f}_{4}$ focuses the
beam for analyzing far field interference.}

\label{im:schemat} 
\end{figure}

The experimental setup is shown in Fig. \ref{im:schemat}. 
We use reflective PLUTO SLM from Holoeye (LCOS type with $1920\times1080$ resolution and 8 $\mu m$ pixel size) and laser with wavelength of 780 $nm$. 
The beam of illuminating laser is enlarged by a Keplerian telescope consisting of the lenses $\textnormal{f}_{1}$ and $\textnormal{f}_{2}$ in order to illuminate the whole SLM surface.
Light reflected and diffracted from the SLM is refocused by $\textnormal{f}_{2}$ lens but only first order diffraction is passed through an iris. 
Lenses $\textnormal{f}_{3}$ and $\textnormal{f}_{4}$ rely far field from iris plane to the Fourier plane camera, while $\textnormal{f}_{2}$
and $\textnormal{f}_{3}$ create SLM image on image plane camera (resolution 1392 $\times$ 1040, pixel size 6.45 $\mu$m, Bassler sca1400-17fm   ).
The half wave plate enables adjusting the splitting ratio between
diagnostic part and further experiment part.

\section{Depth-modulated grating}
\label{sec:grating} Diffraction of beam with arbitrary amplitude
and phase is possible by displaying depth-modulated phase diffraction
grating  $\Psi(x,y)$ on the SLM. The grating is composed of tiles
with depth $a$ and spatial period $d$ and can be described by a formula:
\begin{equation}
\Psi(x,y)=aQ_{d}(x),\quad Q_{d}(x)=2\pi\frac{x}{d}\textnormal{ mod }2\pi.\label{eq:grating}
\end{equation}
Let the grating be illuminated with a plane wave of amplitude $E_{0}$.
Then the electric field corresponding to the first order diffraction off the grating on SLM plane is:
\begin{equation}
E_{1}(x,y)=E_{0}\textnormal{ sinc}(\pi(1-a))e^{-i\pi(1-a)}e^{i\frac{2\pi}{d}x}
\end{equation}

\begin{figure}[h]
\centering \includegraphics[width=1\linewidth]{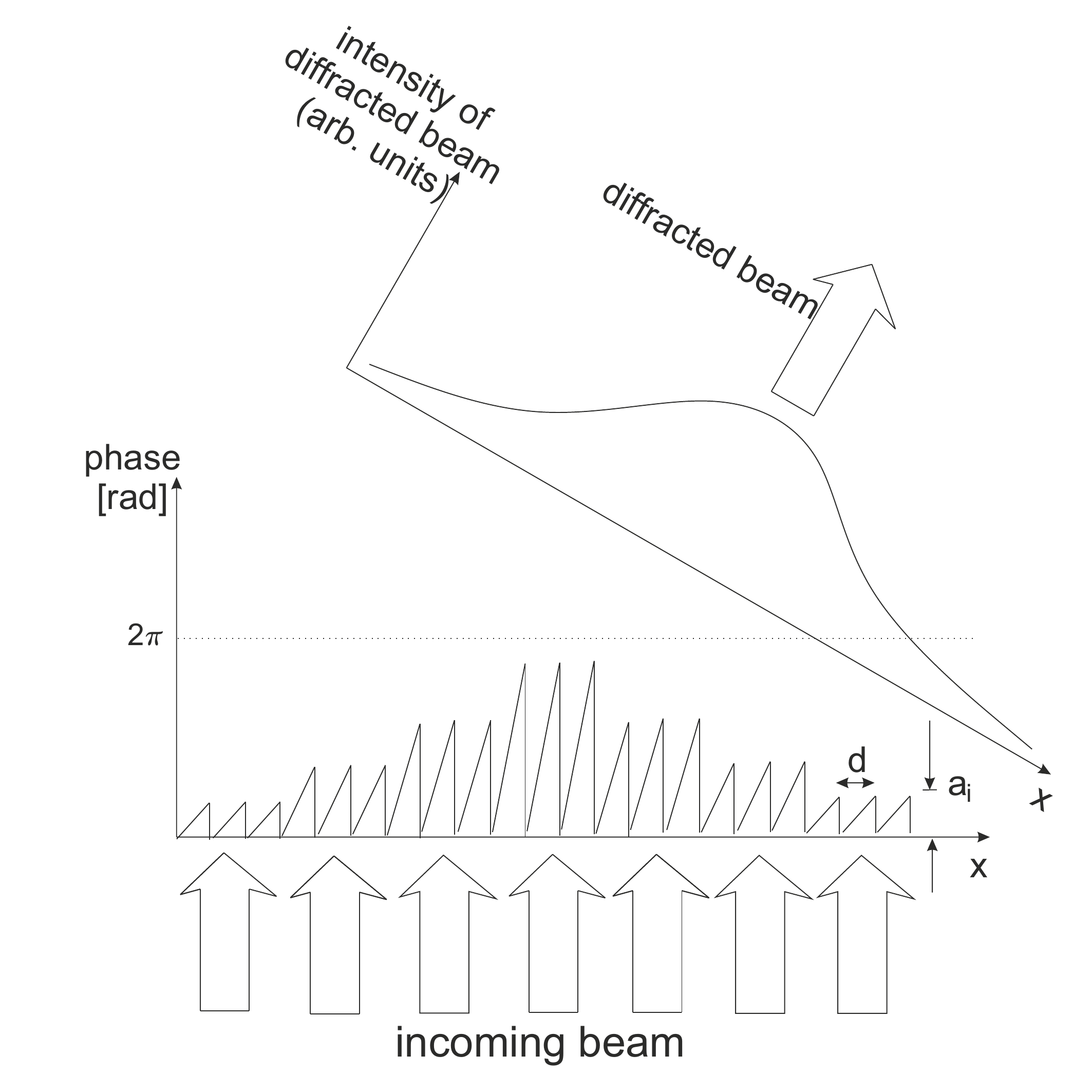} 
\caption{Cross section through a variable-depth phase grating $\Psi(x,y)$ used to diffract beam with arbitrary amplitude. 
It consists of many fragments  with different depths $a_i$.}
\label{im:phase_mask} 
\end{figure}

By adjusting the grating depth $a$ and shifting it laterally
the amplitude and phase of the diffracted beam can be changed at will.
By composing many short fragments of varying depths next to one another,
as illustrated in Fig. \ref{im:phase_mask} beam with arbitrary amplitude
distribution can be diffracted \cite{grating1,grating2}. It is necessary
for the amplitude of desired diffracted beam to change slowly compared to spatial period of the grating $d$ 
so that various orders of diffraction can be separated by spatial filtering.

\begin{figure}[!h]
\centering \includegraphics[width=1\linewidth]{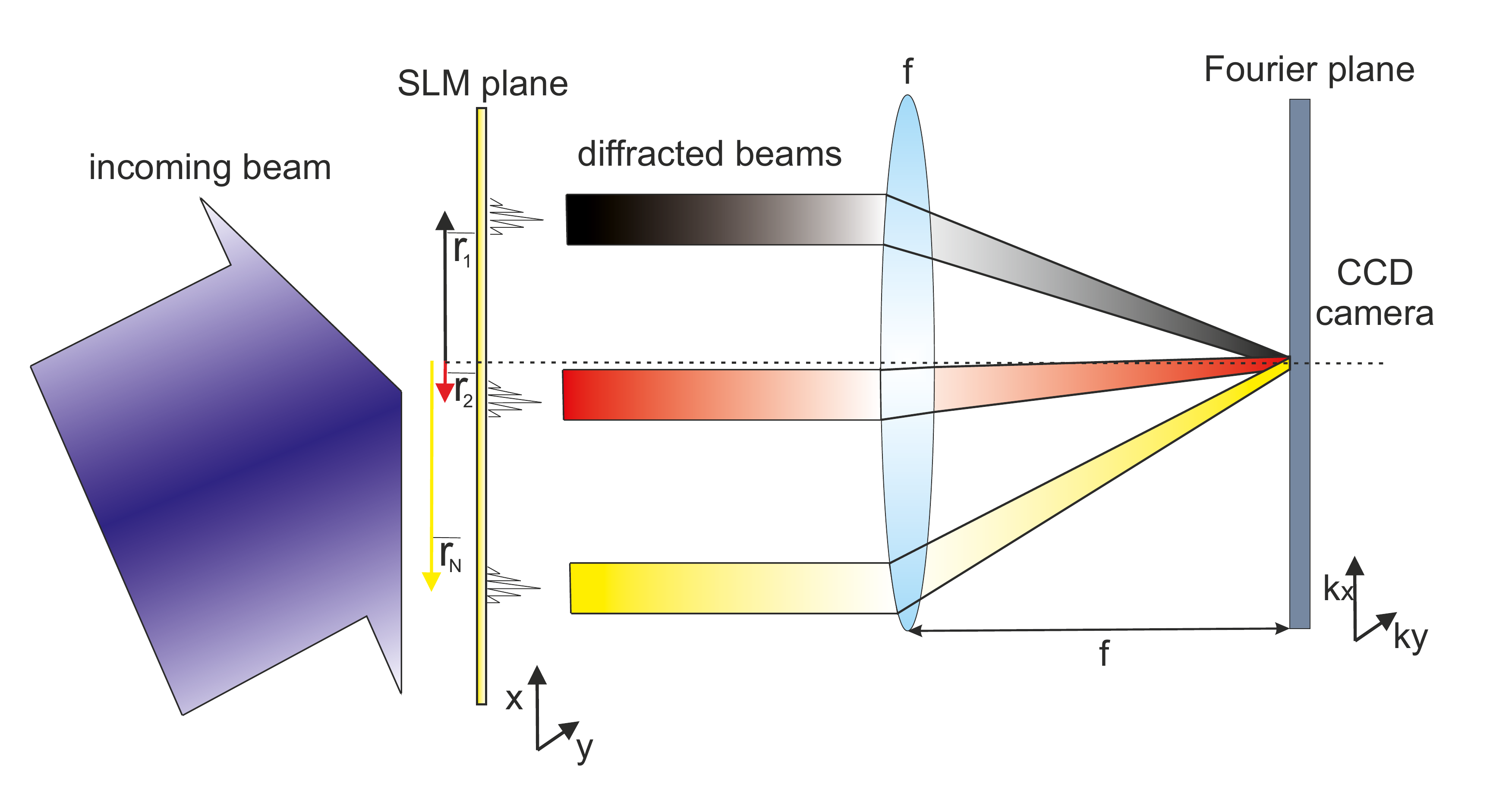} 
\caption{
Concept of measurement used to retrieve the wavefront distortion. 
Several parallel beams diffracted off the SLM around points $\vec r_i$ are focused together and interfere on the camera placed in the Fourier plane.
} \label{im:skupienie} 
\end{figure}

\section{Phase calibration} \label{sec:phase} 
To create a map of SLM phase distortion $\phi(\vec r)$ we measure
relative phases between several parallel beams diffracted off the
SLM by interfering them on a Fourier plane camera as shown in Fig.
\ref{im:skupienie}.

\begin{figure*}[th]
\centering
\begin{subfigure}[b]{0.49\textwidth}
\includegraphics[width=1\linewidth]{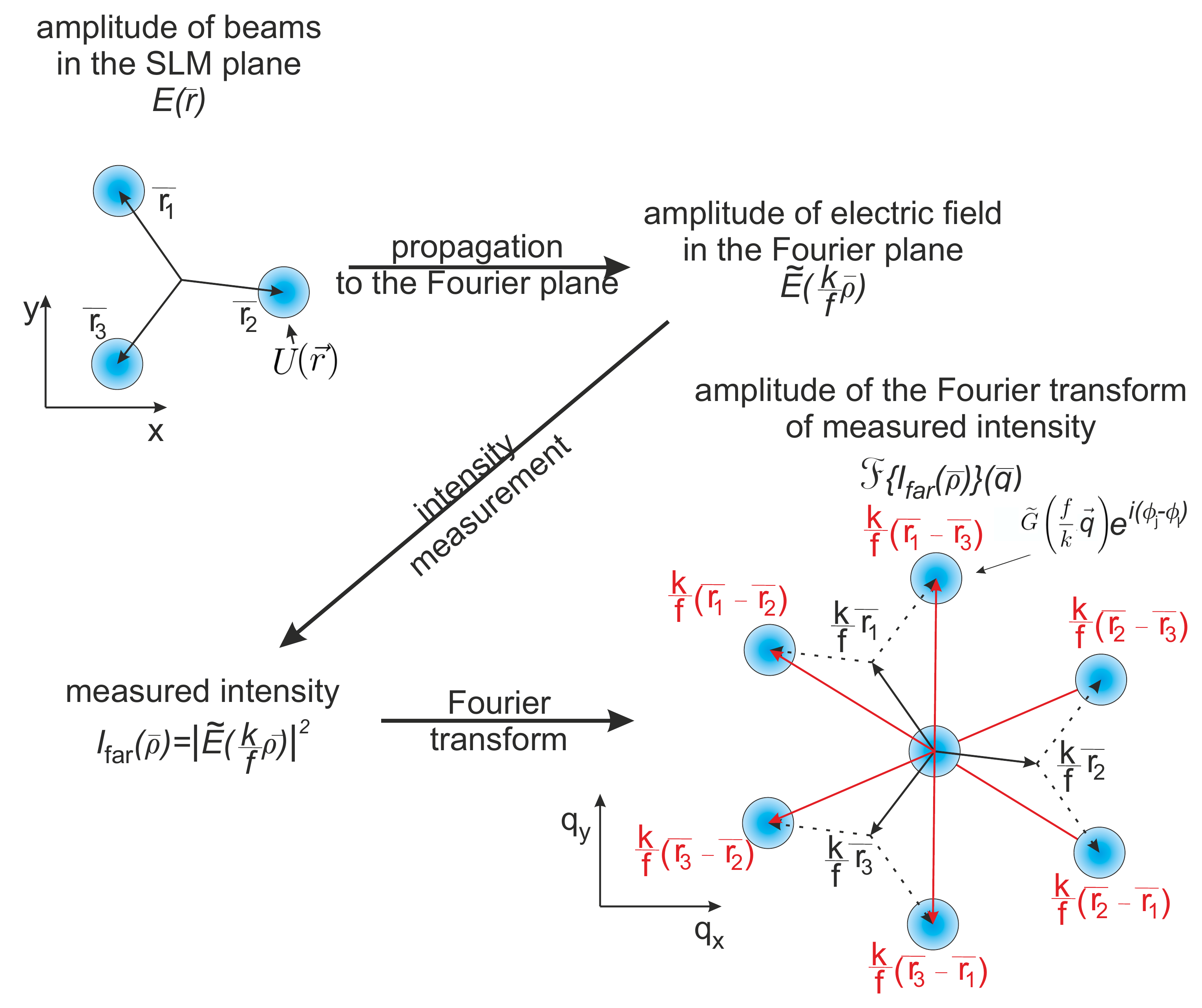}
\caption{}
\end{subfigure}
\begin{subfigure}[b]{0.49\textwidth}
\includegraphics[width=1\linewidth]{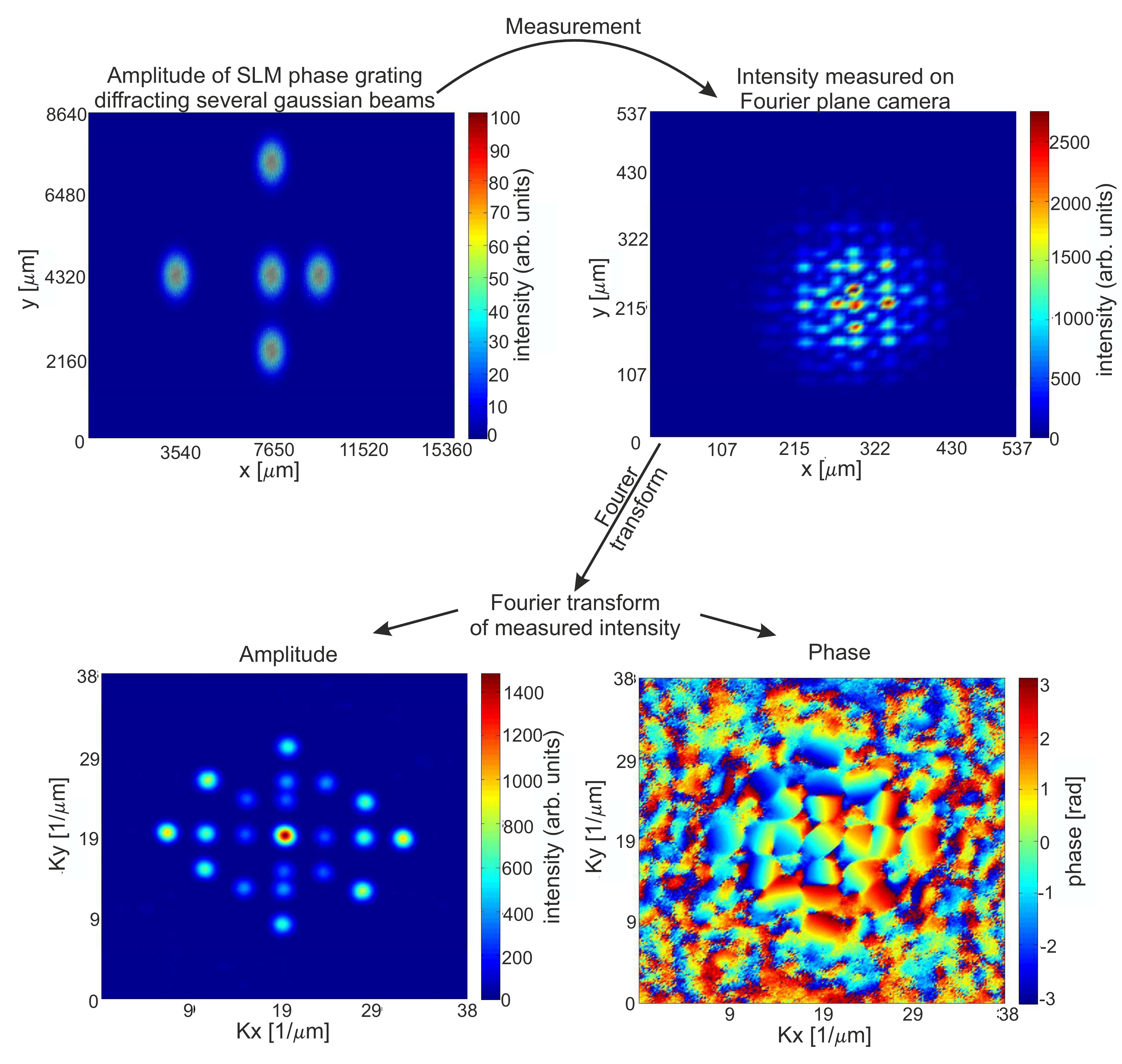}
\caption{}
\end{subfigure}
\caption{ (a) Schematic illustration of the subsequent steps required to measure
relative phases of the beams diffracted off the SLM. (b) Phase grating
displayed on SLM $\Psi(x,y)$, measured intensity in the Fourier plane
$I_\text{far}(\vec{\rho})$ and its Fourier transform $\mathcal{F}\{I_\text{far}(\vec{\rho})\}(\vec{q})$  (amplitude and phase).
}\label{im:fourier2} 
\end{figure*}

The beams are diffracted around different points
$\vec{r_{j}}=(x_i,y_i)$ and focus together on the far field camera. We assume
all beams have same amplitude profile $U(\vec{r})$ on SLM, but may
have different intensities $\alpha_{j}^{2}$ due to nonuniform illumination.
The phases $\phi_{j}=\phi(\vec{r}_{j})$ of the beams are to be measured.
 The amplitude of the beam number $j$ in the SLM plane is therefore
$\alpha_{j}U(\vec{r}-\vec{r_{j}})e^{i\phi_{j}}$.
The intensity in the Fourier plane $I_{\mathrm{far}}(\vec{\rho})$ is proportional
to the modulus square of the Fourier transform of the amplitude on
the SLM plane. Spatial coordinates on the far field camera $\vec\rho$ correspond 
$k_x$ and $k_y$ components of the wavevectors of the Fourier decomposition of the field on the SLM 
through the relation $\vec k=k \vec \rho/f$ where $k=2\pi/\lambda$ is the wavevector of light
and $f$ is the focal length of the Fourier transforming lens.  
Denoting Fourier transform of a function $f(\vec{r})$ in $\vec k$-space by $\mathcal{F}\{f(\vec{r})\}(\vec{k})$
we can write:
\begin{eqnarray}
I_\text{far}(\vec{\rho}) & \propto & \left|\mathcal{F}\left\{\sum_{j=1}^{N}\alpha_{j}U(\vec{r}-\vec{r_{j}})e^{i\phi_{j}}\right\}\left(\frac{k}{f}\vec{\rho}\right)\right|^{2}\nonumber \\
 & = & \left|\sum_{j=1}^{N}\alpha_{j}\tilde{U}\left(\frac{k}{f}\vec{\rho}\right)e^{-i\frac{k}{f}\vec{\rho}\vec{r_{j}}+i\phi_{j}}\right|^{2}\nonumber \\
 & = & 
\left|\tilde{U}\left(\frac{k}{f}\vec{\rho}\right)\right|^{2} 
\sum_{j,l} e^{i\frac{k}{f}(\vec{r_{l}}-\vec{r_{j}})\vec{\rho}}
\alpha_{j}\alpha_{l}e^{i(\phi_{j}-\phi_{l})}\label{eq:intensity}
\end{eqnarray}
above $N$ is the number of beams 
and $\tilde{U}(\vec{k})$ denotes Fourier transform of single beam amplitude profile,  $\tilde{U}(\vec{k})=\mathcal{F}\{U(\vec{r})\}(\vec{k)}$. 
The intensity of the interference pattern is the Fourier plane $I_{\mathrm{far}}(\vec{\rho})$ is a sum of contributions from each possible pair of beams. 
They overlap to form a fringe modulation under the envelope set by the far field image of any single beam $\left|\tilde{U}\left({k}\vec{\rho}/f\right)\right|^{2} $. 

The crucial observation is contributions to the interference pattern $I_\text{far}(\vec{\rho})$ coming from each pair of beams 
can be separated by Fourier transforming the $I_\text{far}(\vec{\rho})$ thus revealing the phase differences $\phi_{j}-\phi_{l}$.
This step is performed numerically on a PC and transforms from the far-field coordinates $\vec\rho$ into its conjugate coordinates $\vec q$:
\begin{equation}
\mathcal{F}\{I_\text{far}(\vec{\rho})\}(\vec{q})
\propto
\sum_{j,l}G\left(\frac{f}{k}\Bigl(\vec{q}-\frac{k}{f}(\vec{r_{l}}-\vec{r_{j}})\Bigr)\right)\alpha_{j}\alpha_{l}e^{i(\phi_{j}-\phi_{l})}\label{eq:fourier}
\end{equation}
where $G(\vec{q})=\mathcal{F}\left\{|\tilde{U}(\vec{k})|^{2}\right\}(\vec{q)}$
depends only on the amplitude profile $U(\vec{r})$ of each beam.
The value of $G(0)$ has to be real, because it is Fourier transform of a real function. 
The amplitude of the Fourier transform $\mathcal{F}\{I_\text{far}(\vec{\rho})\}(\vec{q})$
has maxima at locations $\vec q_{l,j}=(\vec{r_{l}}-\vec{r_{j}})k/f$. We
assume they are separated, that is each difference of the vectors
$\vec{r_{l}}-\vec{r_{j}}$ is unique. Then from the formula (\ref{eq:fourier})
it follows that the phase at a point $\vec q_{l,j}=(\vec{r_{l}}-\vec{r_{j}})k/f$
equals the phase difference between respective beams $\phi_{j}-\phi_{l}$.
The above steps are illustrated in Fig.~\ref{im:fourier2}.

Measurement with $N$ beams yields ${N(N-1)}/{2}$ phase
differences, that is ${N(N-1)}/{2}$ equations for $N$ unknown
phases. We find phases by least square fit. 

\begin{figure}[h]
\centering \includegraphics[width=1\linewidth]{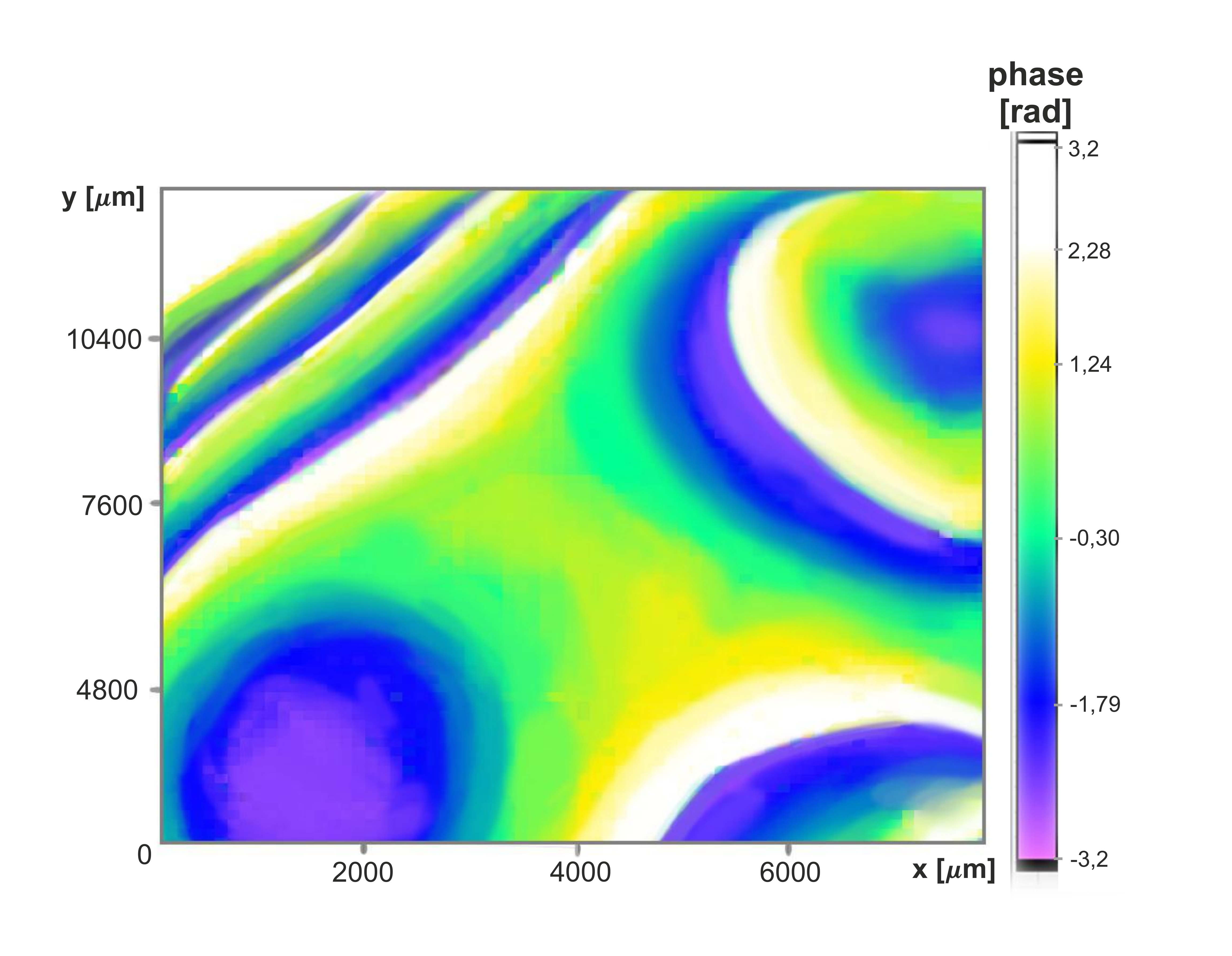} 
\caption{Phase distortion map $\phi(\vec{r})$, created
by averaging results of 5-point measurements from 500 images and 4-point
measurements from 200 images. This is a combined effect of the SLM,
illuminating beam wavefront distortion and focusing optics.
}\label{im:faza} 
\end{figure}

The number of beams $N$ is limited because the vectors
$\vec{r_{l}}-\vec{r}_{j}$ have to be unique. For this reason to make
phase distortion map $\phi(\vec{r})$ we perform a series of measurements.
We diffract $N=5$ beams, which is good tradeoff between complexity and precision. First beam is in the middle of the
SLM $\vec{r}_{1}=0$ and serves as a common phase reference i.e. we
take $\phi(0)=0$. Other beams are placed around central beam and
shifted after acquiring each far field image so as to cover entire
surface of the SLM except near the centre. To determine the phase
near the centre we perform additional measurements. We choose three reference
points at the periphery of the SLM and shift central beam around the
middle of the SLM. By averaging all the results we retrieve the phase
distortion map $\phi(\vec{r})$ presented in Fig.~\ref{im:faza}.
The entire measurement consists in acquiring 700 images and takes
about 5 minutes which is limited predominantly by delays in simple LabVIEW software while changing phase masks on the SLM and acquisition from the camera.

Note the scheme of calibration described above compensates for the
wavefront distortions of the optics between the SLM and a certain
calibration plane which is imaged on one camera and Fourier transformed
to the other camera. For proper operation of the scheme the optical setup
has to fulfil the following conditions. The SLM has to be imaged with little
blur, of the order of several periods of phase grating $d$ so that
the desired final spatial amplitude variations at relied with sufficient
fidelity. The far field camera has to register true Fourier transform of the calibration
plane field. The fidelity has to be high at small angles corresponding
to angular spread of probing Gaussian beams and the resolution good
enough to resolve fringes due to their interference. 

In our setup (depicted in Fig. \ref{im:schemat}) the calibration plane
is located at the SLM image camera that is 79 mm behind the center of the
PBS and the image there is magnified by a factor of $-0.7$. 
Due to particular lens arrangement the wavefront there is diverging
with $1059$~mm radius of curvature when the entire beam is focused
in the Fourier plane. Thus to produce certain wavefront $\theta(\vec{r})$
at physical location of our image plane one has to program spatial
phase $\theta(\vec{r})+r^{2}/(R\lambda)$ with $R=2125$~mm 
due to magnification with respect to image plane.

\begin{figure}[!h]
\centering
\includegraphics[width=0.7\linewidth]{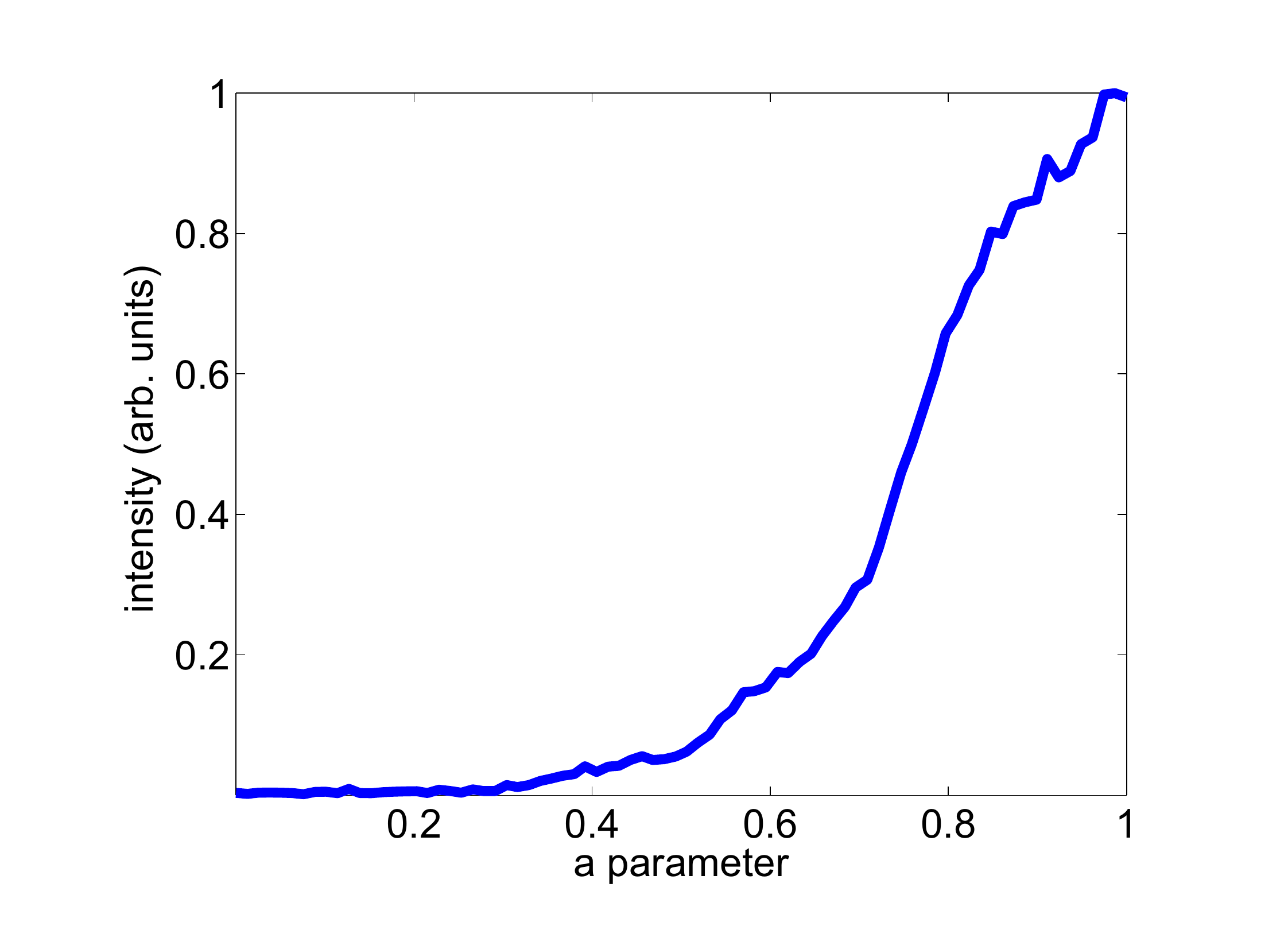} 
\caption{Measured intensity of diffracted wave $\alpha(a)^{2}$ as a function of phase grating depth $2\pi a$ normalized to unity.}
\label{im:alpha_a}
\end{figure}

\section{Amplitude correction} \label{sec:amplitude}

\begin{figure}[!h]
\centering 
\begin{subfigure}[b]{0.23\textwidth} 
	\includegraphics[width=1\textwidth]{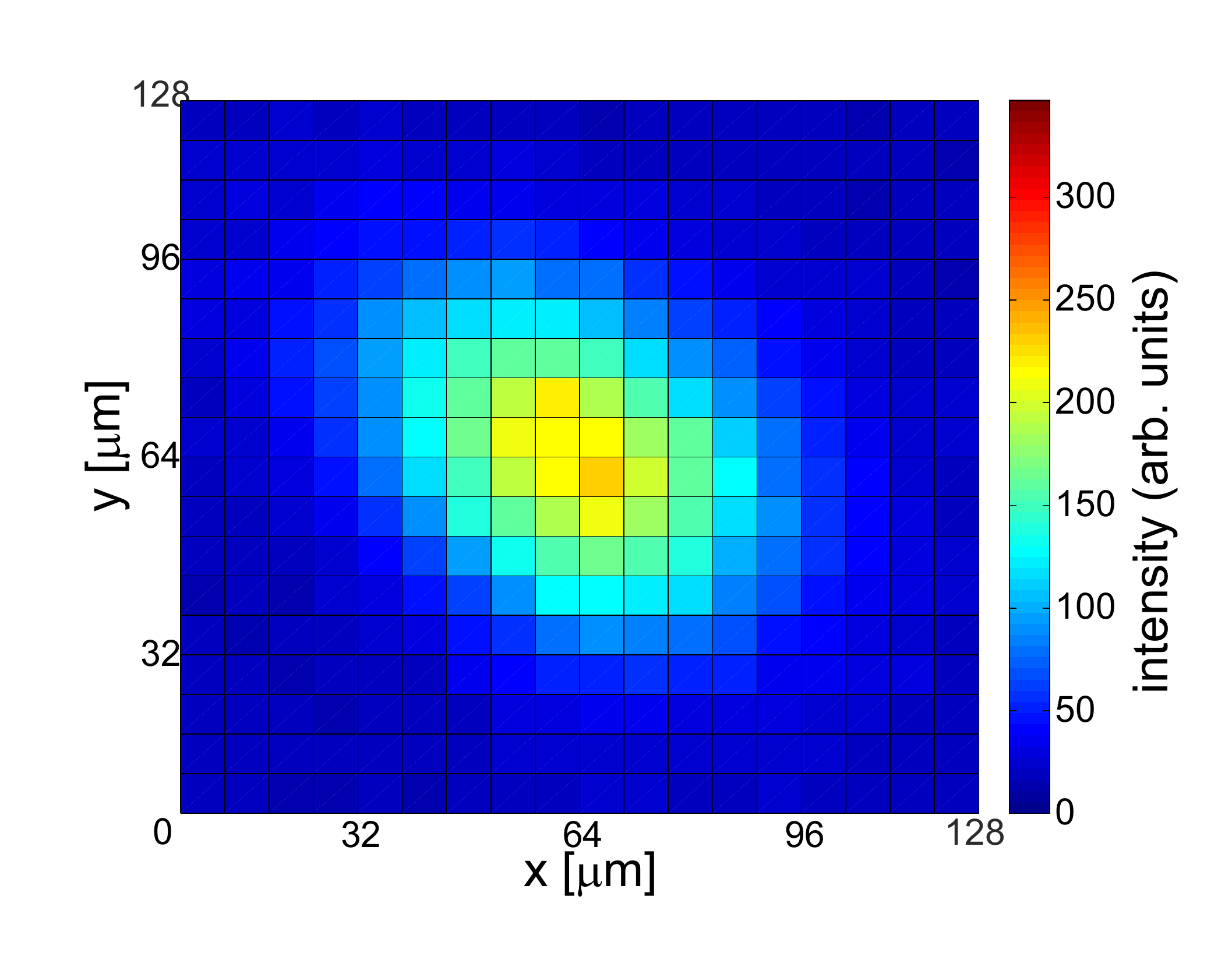}
	\caption{} 	\label{im:gauss_przed}  
\end{subfigure} \begin{subfigure}[b]{0.23\textwidth}
	\includegraphics[width=1\textwidth]{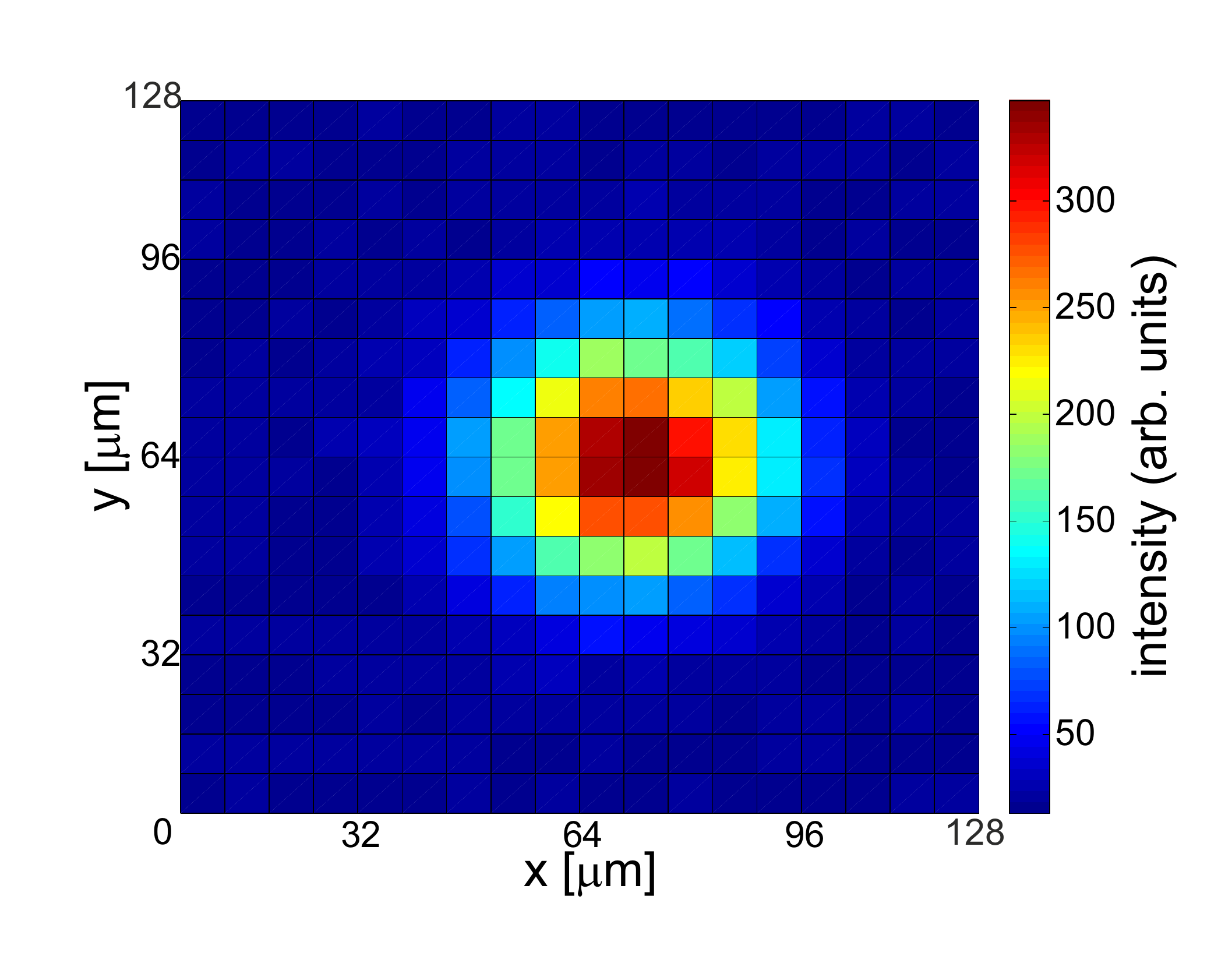} 
	\caption{}	\label{im:gauss_po} 
\end{subfigure} 
\caption{Intensity of Gaussian beam with 1.5~mm diameter on SLM focused on Fourier plane
camera (a) before phase correction (b) after phase correction.}
\label{im:gauss} 
\end{figure}

\begin{figure}[!h]
\centering \begin{subfigure}[b]{0.49\linewidth} \includegraphics[width=1\textwidth]{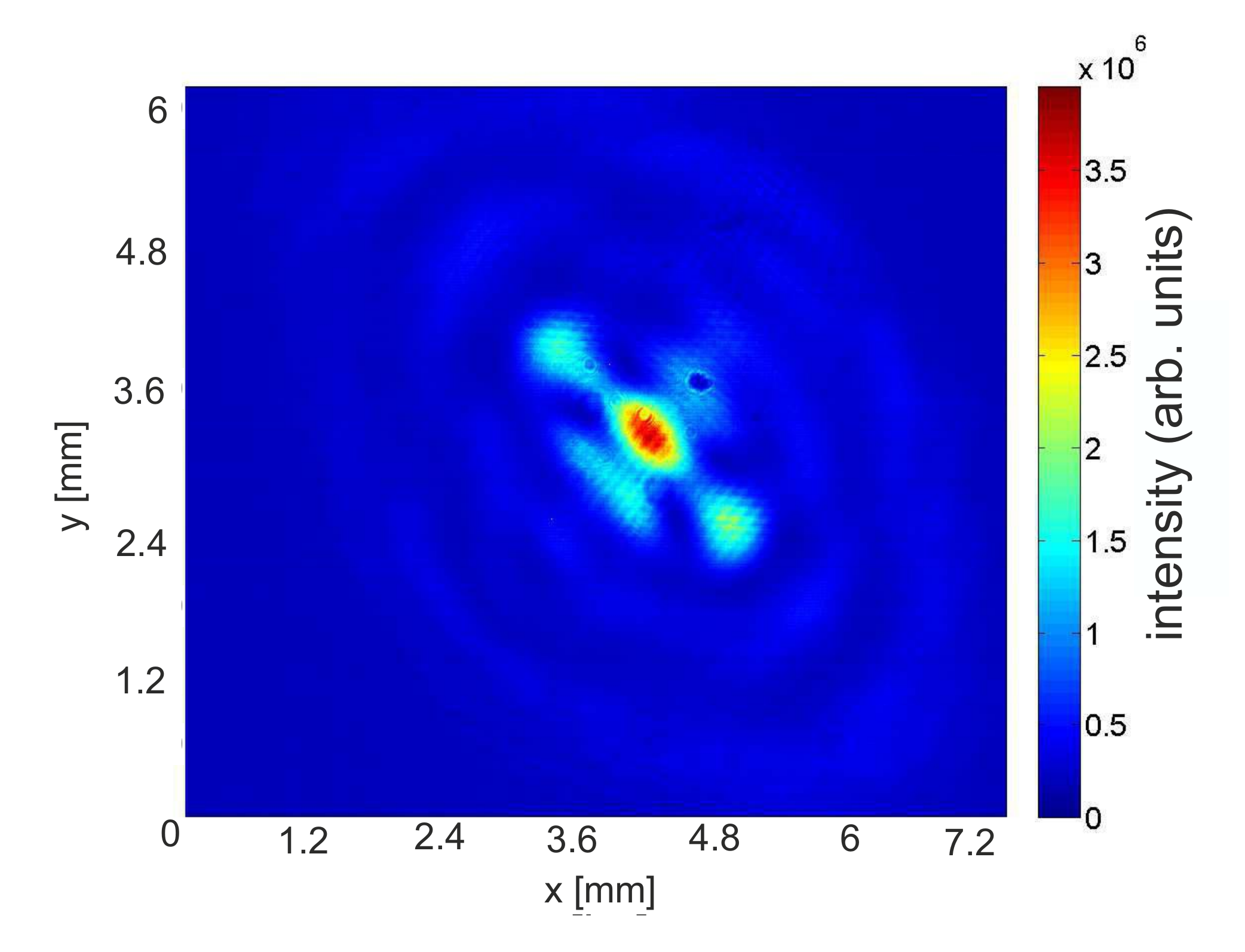}
\caption{}
\label{im:bessel_przed} \end{subfigure} 
\begin{subfigure}[b]{0.49\linewidth}
\includegraphics[width=1\linewidth]{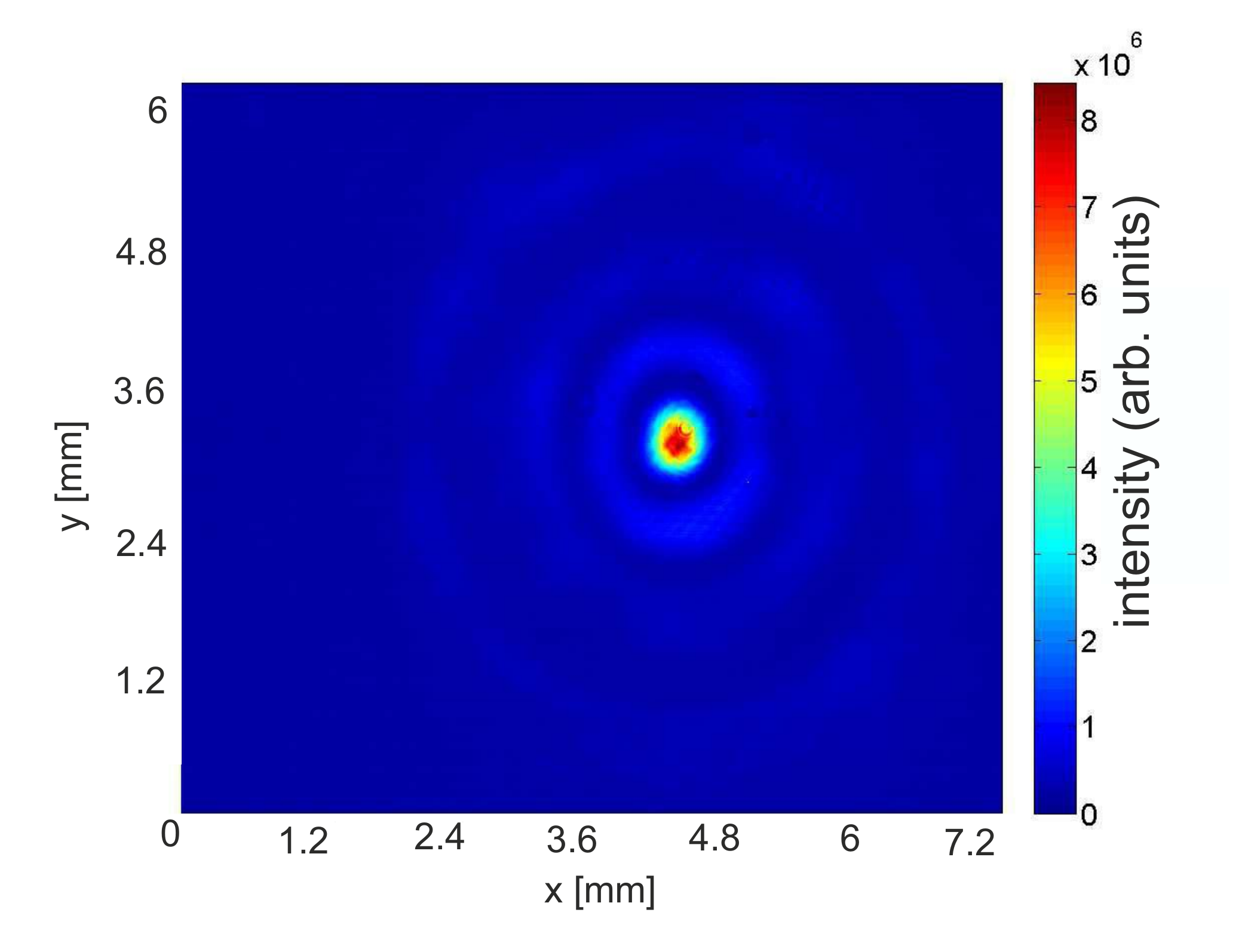} \caption{}
\label{im:bessel_po} \end{subfigure}
\begin{subfigure}[b]{0.49\linewidth}
\includegraphics[width=1\linewidth]{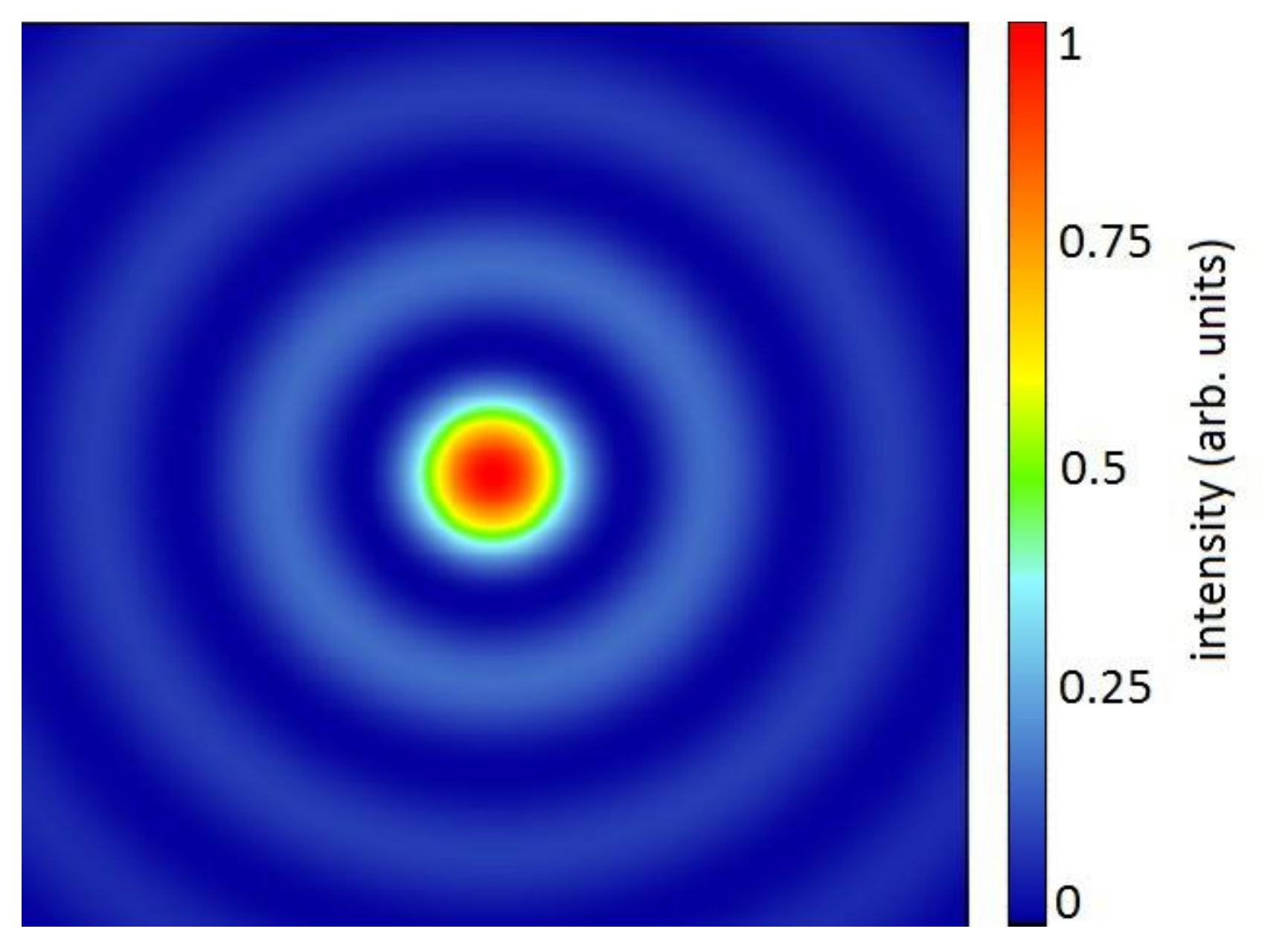} \caption{}
\label{im:bessel_zero} \end{subfigure}
 \caption{Intensity of zero Bessel beam diffracted off the SLM after two meters
propagation (a) before phase and amplitude correction (b) after phase
and amplitude correction. Figure (c) presents theoretical intensity of zero Bessel beam. }
\label{im:bessel}
\end{figure}
 
The amplitude of the beam which illuminates
the SLM has certain nonuniform distribution $J(\vec{r})$. It can
be measured by displaying phase grating with spatially constant depth
equal $2\pi$ (see Fig.~\ref{im:phase_mask}) on the whole SLM surface
and registering an intensity distribution on image plane camera. Having
measured $J(\vec{r})$ the amplitude of the beam to be diffracted
can be multiplied by a factor ${\sqrt{\max\,J(\vec{r})/J(\vec{r})}}$
to compensate the effects of nonuniform illumination. 

Next the diffraction efficiency versus depth of the phase grating
$2\pi a$ is measured. The resulting $\alpha(a)^{2}$ normalized to
unity is shown in Fig.~\ref{im:alpha_a}. The inverse of $\alpha(a)$
function $M(\alpha)$ informs us that to diffract wave with arbitrary amplitude
$\alpha$  the phase grating of depth  $a=M(\alpha)$ has to be used. 
Correcting for nonuniform illumination the required phase grating depth at point $\vec r$ becomes
$a=M\left[\alpha(\vec r) {\sqrt{\max\,J(\vec{r})/J(\vec{r})}} \right]$.

\section{Calibrated programming of SLM} \label{sec:verify} 

\begin{figure*}[!ht]
\centering
\begin{subfigure}[b]{0.32\textwidth} \includegraphics[width=1\textwidth]{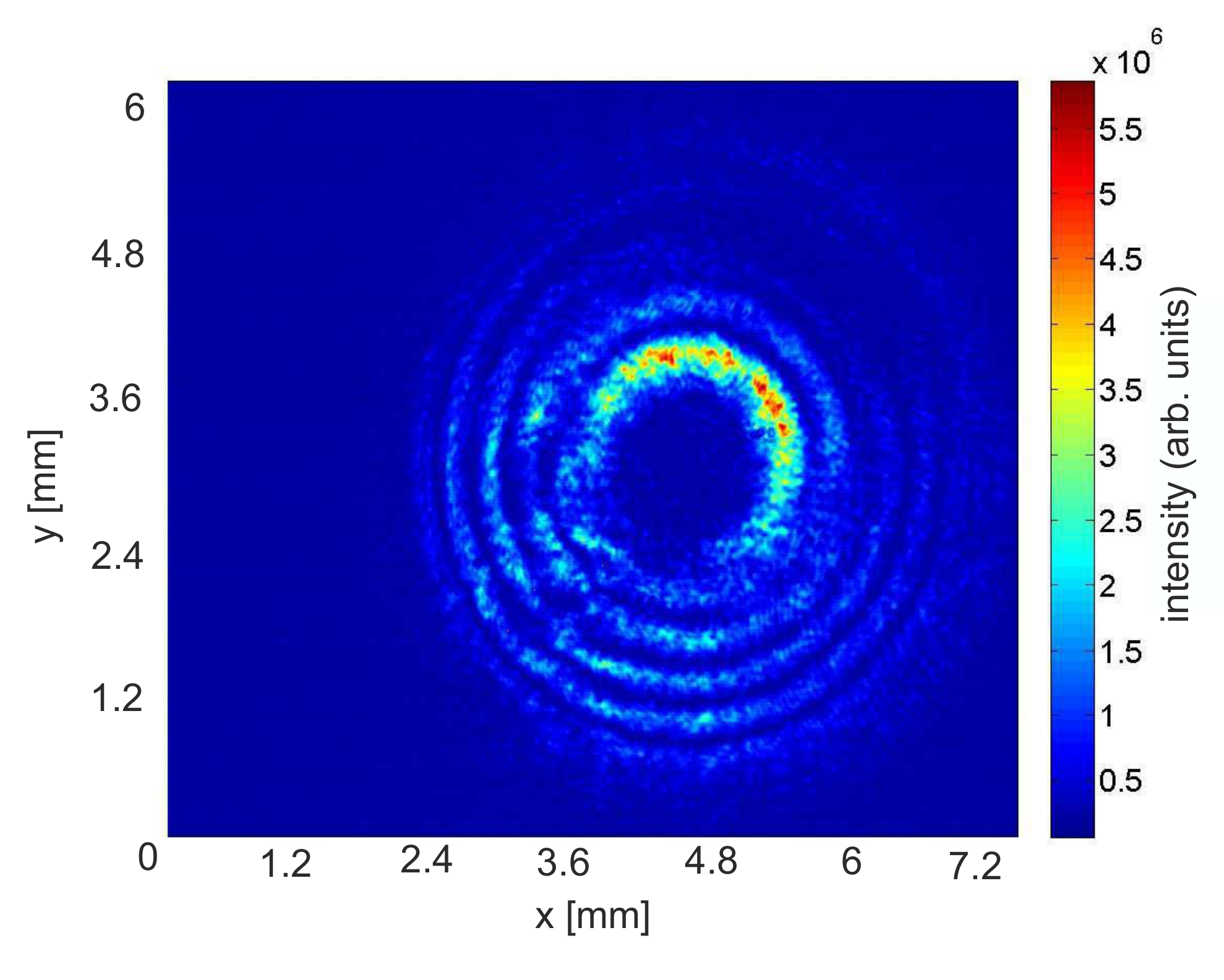}
\label{im:bessel7+8_1} 
\caption{}
\end{subfigure} 
\begin{subfigure}[b]{0.32\textwidth} \includegraphics[width=1\textwidth]{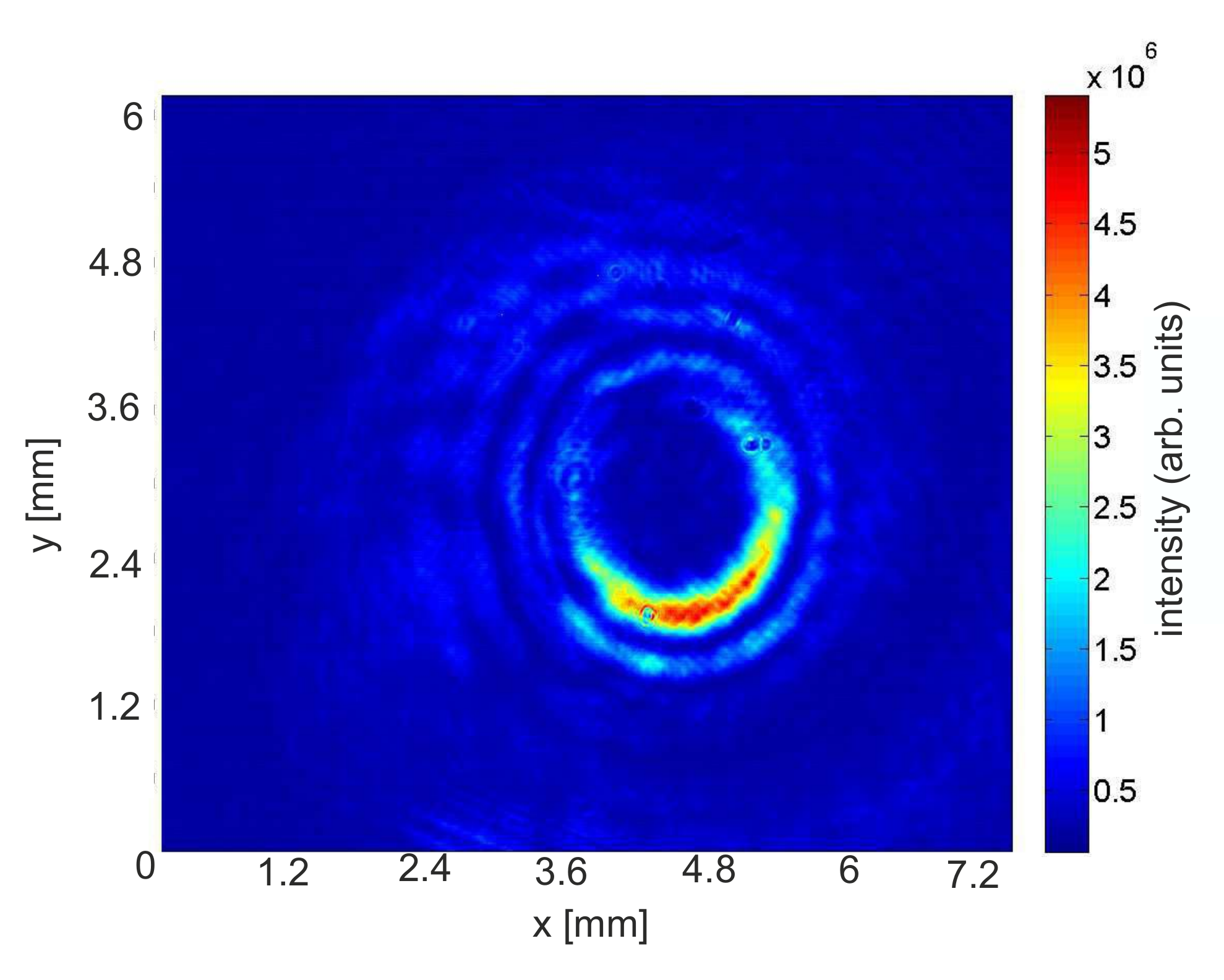}
\label{im:bessel7+8_2}
\caption{}
\end{subfigure} 
\begin{subfigure}[b]{0.32\textwidth} \includegraphics[width=1\textwidth]{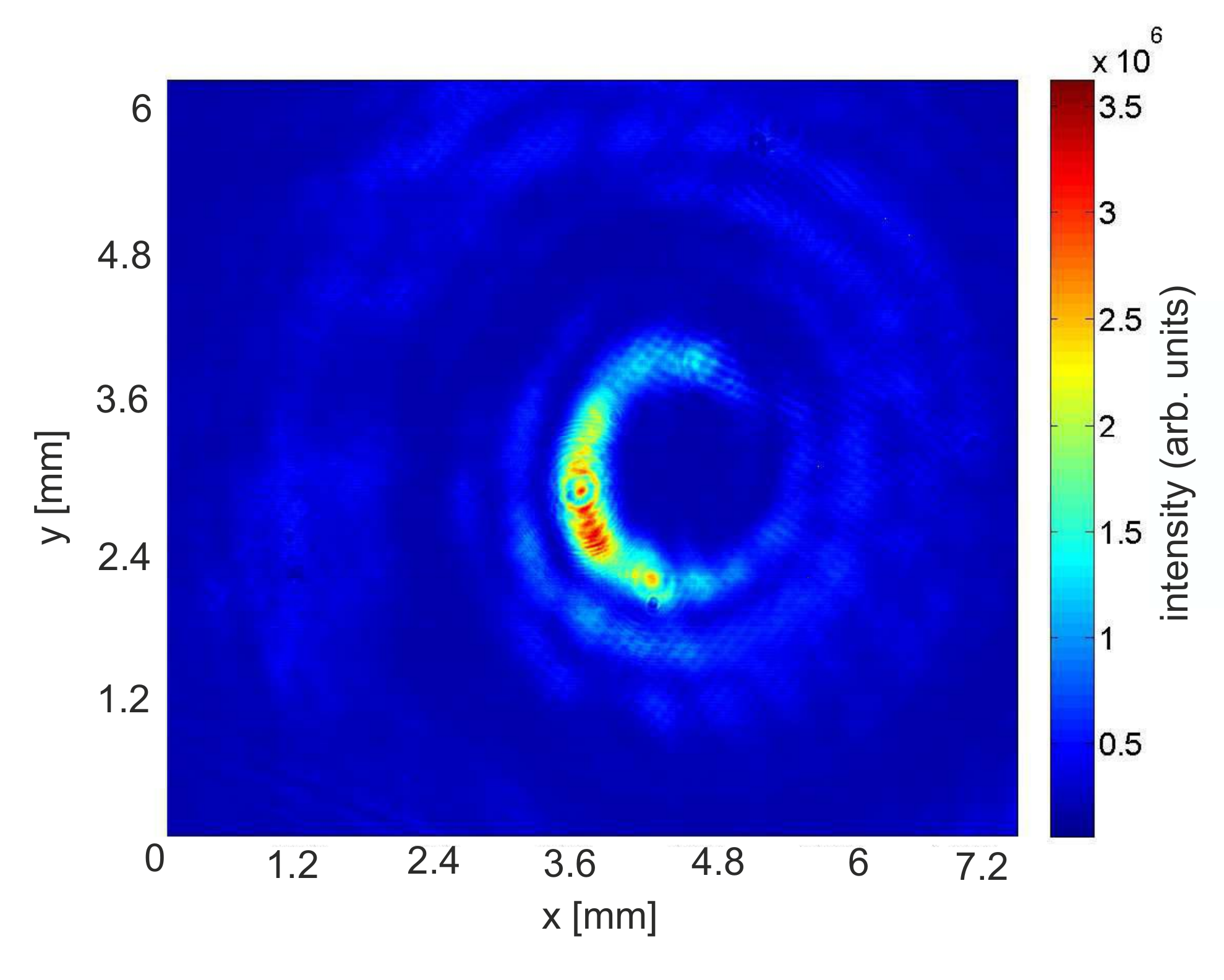}
\label{im:bessel7+8_3} 
\caption{}
\end{subfigure} 
\begin{subfigure}[b]{0.32\textwidth} \includegraphics[width=1\textwidth]{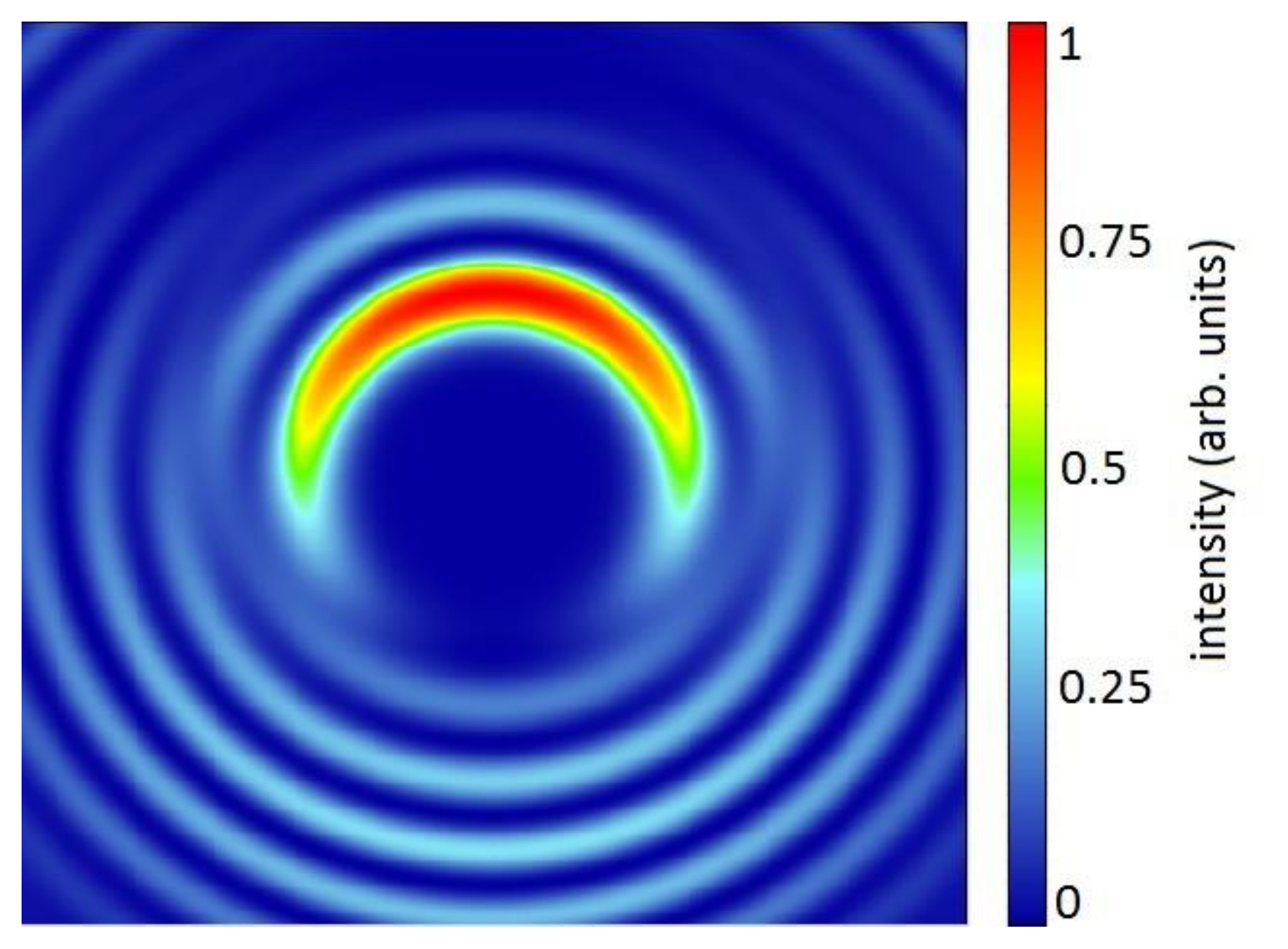}
\label{im:bessel7+8-22}
\caption{}
\end{subfigure}
\begin{subfigure}[b]{0.32\textwidth} \includegraphics[width=1\textwidth]{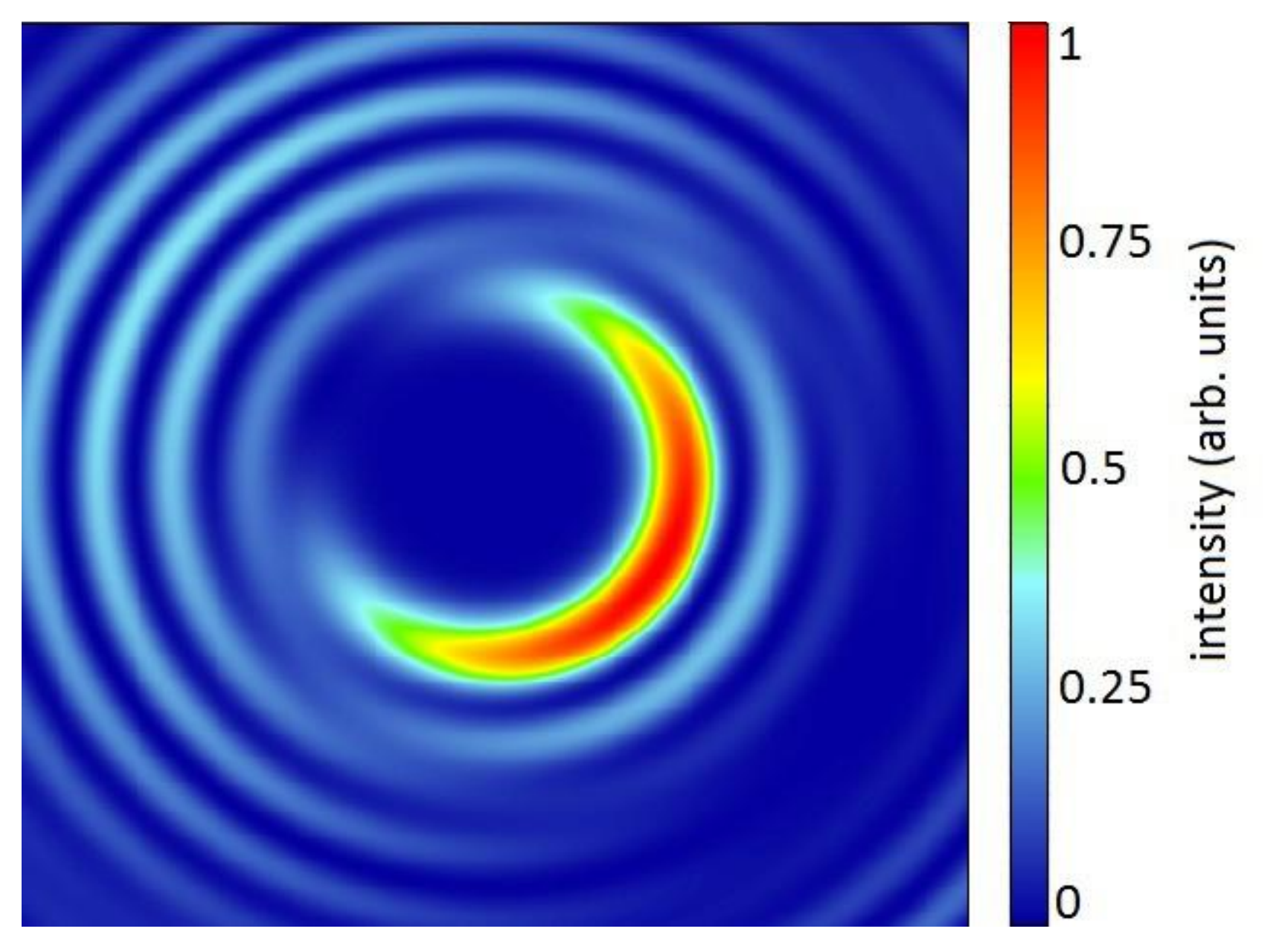}
\label{im:bessel7+8-33}
\caption{}
 \end{subfigure}
\begin{subfigure}[b]{0.32\textwidth} \includegraphics[width=1\textwidth]{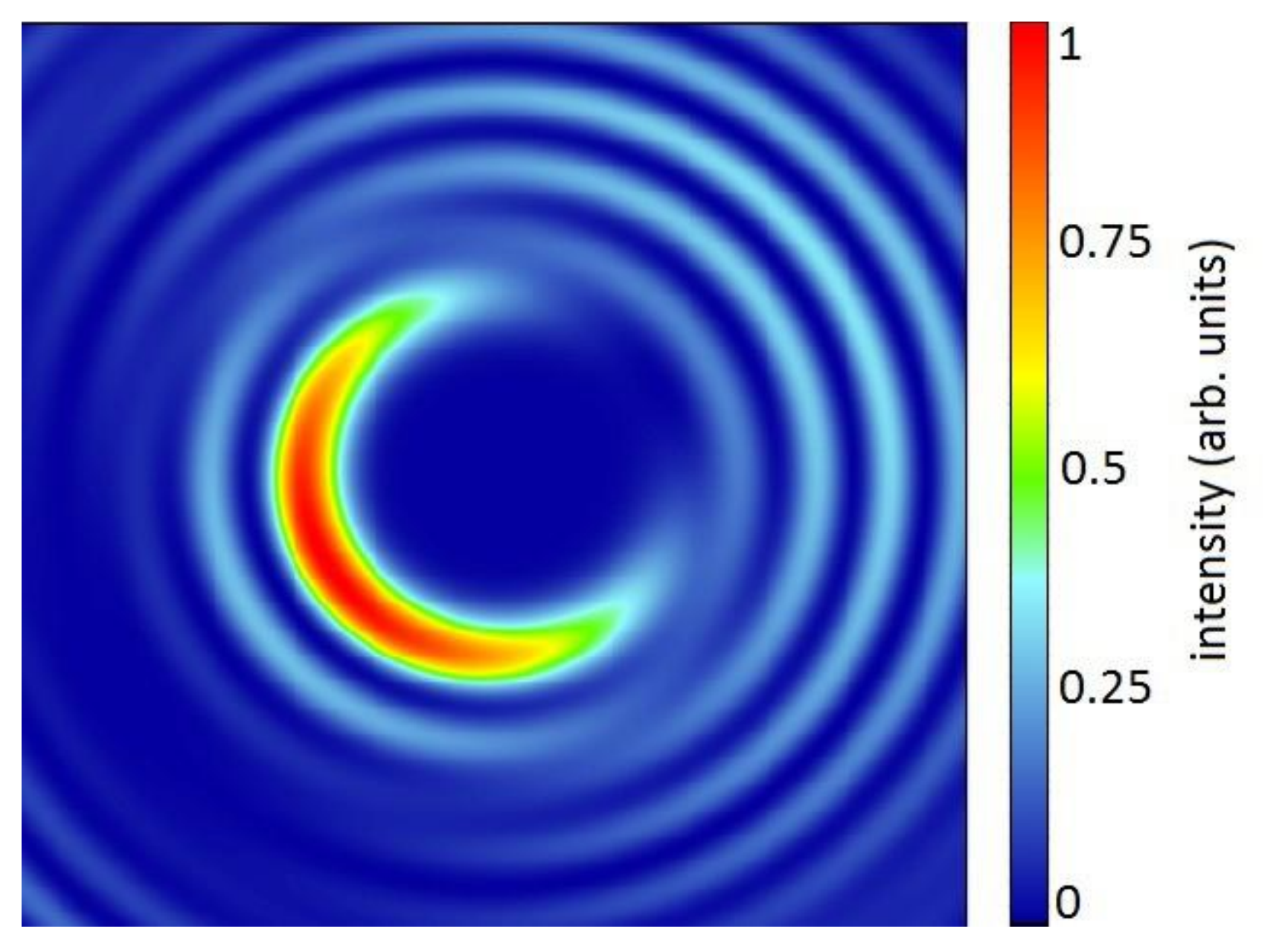}
\label{im:bessel7+8-44} 
\caption{}
\end{subfigure}

\caption{Intensity of seventh and eighth Bessel beam superposition
(a) in the image plane of SLM and after propagation by (b) 75~cm or (c) 150~cm. Figures (d), (e) and (f) shows the simulation result.}
\label{im:bessel7+8_prop} 
\end{figure*}

Here we formulate a recipe for diffracting beam of any amplitude and phase profile described by function $A(\vec{r})e^{i\theta(\vec{r})}$.
Suppose that $Q_{d}(\vec{r})$ is elementary grating defined in Eq.~(\ref{eq:grating}),
$\phi(x)$ is a SLM phase distortion (see Fig. \ref{im:faza}),
${\sqrt{\max\,J(\vec{r})/J(\vec{r})}}$ is amplitude
correction factor and M($\alpha$) is inverse of diffraction efficiency
function. Then to diffract the desired beam we project the following
phase grating on the SLM:
\begin{eqnarray}
\Psi(\vec{r}) & = & 
a(\vec r) Q_{d}
\left(x
+d\frac{\theta(\vec r)-\phi(\vec r)}{2\pi}
-\frac{d}{2} a(\vec r),
y  \right),
\\
a(\vec r) &=&M\left(\frac{\sqrt{\max\,J(\vec{r})}}{\sqrt{J(\vec{r})}}A(\vec r)\right).
\end{eqnarray}
The phase of grating $Q_{d}(\vec r)$  is shifted to compensate precalibrated
phase distortion $\phi(\vec r)$ and amplitude dependent phase to produce
desired wavefront $\theta(\vec r)$ and amplitude $A(\vec r)$.\\

As a first test we check Gaussian beam focusing.
Gaussian beam with flat wavefront minimizes spot size in Fourier plane, so
near perfect focusing confirms calibration. Fig.~\ref{im:gauss}
compares measured far field intensity for a Gaussian beam diffracted
off the SLM before and after calibration. Diffracted
beam diameter is 1.5~mm and laser wavelength is 780~nm, so diameter
of ideal Gaussian beam on the camera should be about 26~$\mu$m. 
We find focal plane beam diameter of 28.5~$\mu$m by fitting. 

A less trivial test consists in creating Bessel beams and
verifying their propagation properties. Bessel beams are solutions
of wave equation which are diffraction resistant \cite{bessel}. 
The family of Bessel beams amplitude are expressed in cylindrical coordinates by: 
\begin{equation}
B_{n}(r,\phi,z)=J_{n}(\frac{r}{w})e^{in\phi}e^{ik_{z}z}\label{eq:bessel}
\end{equation}
\begin{equation}
k_{z}=\sqrt{\frac{\omega^{2}}{c^{2}}-\frac{1}{w^{2}}}\label{eq:k_ro_k_z}
\end{equation}
where $J_n$ is n-order Bessel function, $w$ is the beam radius, $\frac{\omega}{2\pi}$
is frequency of light and $c$ is the speed of light.

Initially we created a 0th Bessel beam with $w=0.15\mathrm{mm}$ and
verified its diffraction properties. Results are presented in Fig. \ref{im:bessel}.

Finally we created a superposition of 7th and 8th order Bessel beams. 
Due to different azimuthal phases the sum of these two beams looks
like crescent \cite{key-1} as depicted inf Fig. \ref{im:bessel7+8_prop}. 
Moreover Eq. (\ref{eq:k_ro_k_z}) rules that by setting
different diameters $w$ of the component beams we can mismatch their phase
velocities. This results in rotation of the crescent profile around
beam axis along propagation. In Fig.~\ref{im:bessel7+8_prop} we
present measured intensity of beam superposition at three
distances: 0 cm, 75 cm and 150 cm from the PBS.


In addition to the tests described above we characterized the beam outgoing from our system with commercial HASO 32 Shack-Hartmann wavefront sensor. We program our modulator to diffract uniform rectangular beam.
We compared the wavefronts with and without an additional lens inserted right in front of the laser (see Fig. \ref{im:schemat}) which imprinted curvature onto the wavefront illuminating the SLM. 
Using Shack-Hartmann sensor we retrieved an extra wavefront curvature
corresponding to R=5.5 $\pm$ 0.1 m in the SLM plane, with error due to
uncertanities in relay lens positions. With our method we measured R=
5.3 m. The difference between those two amounts to $\lambda/4$
wavefront difference at the rim of the SLM.
The RMS deviation of raw measurements from paraboloid has random structure with RMS $\lambda/80$ in case of  Shack-Hartmann sensor and RMS $\lambda/25$ in case of our method. 

When the modulator is programmed to produce flat-wavefront beam in the image plane the Shack-Hartmann sensor measures $\lambda/4$ RMS deviation from flat wavefront over entire beam. 
To understand this result note that the path leading to the Fourier plane camera used for calibration at the path leading to Shack-Hartmann sensor differ by extra reflection from PBS and biconvex singlet lens. 
Thus we attribute this result to the quality of PBS stated as $\lambda/4$ by the manufacturer and spherical aberration of the lens. 
For experiments requiring greater wavefront precision those two elements or their equivalents have to be of better quality.

\section{Conclusions}

We performed proof-of-principle demonstration of a simple method to calibrate and compensate the total wavefront distortion and nonuniform illumination in an SLM setup. 
The method relies on diffracting several Gaussian beams off the SLM and registering their interference pattern on a camera placed in the far field. 
We confirmed the accuracy of calibration by diffracting the Bessel beams and a superposition of two Bessel beams which is rotating during propagation as expected. 
Additionally we verified our method and commercial Haso 32 Shack-Hartmann wavefront sensor. 

We implemented our method using standard singlet lenses and optical components and a pair of digital cameras. 
The method relies on registering position of rather sparse interference fringes thus very basic cameras are sufficient. 
The setup could be even further simplified by performing amplitude calibration without the image plane camera. 
This would be accomplished by diffracting a single small beam off the SLM, scanning its position across the surface and registering the intensity using the Fourier plane camera.

Our method is well suited to compensating wavefront errors introduced by the optical elements used to rely
the SLM onto the final use plane as long as this final plane is faithfully Fourier transformed onto far field camera.
The algorithm presented directly probes phase differences between multiple points. 
Any setup sufficient for registering both near and far field properties of diffracted beams 
is adequate for our method. We therefore believe it could be a natural choice in a number of experiments for both diagnostic and calibration purposes.

\section*{Acknowledgments}
We gratefully acknowledge the support from the National Science Centre (Poland) Grant No. 2011/03/D/ST2/01941. 
We also would like to thank Micha\l{} Parniak and Micha\l{} D\k{a}browski for their hints and Konrad Banaszek for his generous support.

\end{document}